\documentstyle[aps,twocolumn]{revtex}

\include{incpic}

\newcommand{\R}{I\!\!R}
\newcommand{\CC}{{\cal C}}
\newcommand{\II}{{\cal I}}
\newcommand{\OO}{{\cal O}}
\newcommand{\RR}{{\cal R}}

\begin{document}
\title{Probability current tornado loops in three--dimensional
scattering} 
\author{P.~Exner$^{1,2}$ and P.~\v{S}eba$^{1-3}$}

\address {$^1$ Nuclear Physics Institute, Academy of Sciences,
25068 \v{R}e\v{z} near Prague, Czech Republic \\
$^2$ Doppler Institute, Czech Technical University,
B\v{r}ehov\'{a} 7, 11519 Prague, Czech Republic \\
$^3$ Physics Department, Pedagogical University, V\'{\i}ta
Nejedl\'{e}ho 573, 50003 Hradec Kr\'{a}lov\'{e}, Czech Republic \\
{\em exner@ujf.cas.cz, seba@kostelec.czcom.cz}}
\date{\today}
\maketitle

\begin{abstract}
We consider scattering of a three--dimensional particle on a finite
family of $\,\delta\,$ potentials. For some parameter values the
scattering wavenctions exhibit nodal lines in the form of closed
loops, which may touch but do not entangle. The corresponding
probability current forms vortical singularities around these lines;
if the scattered particle is charged, this gives rise to magnetic
flux loops. The conclusions extend to scattering on hard obstacles or
smooth potentials.
\end{abstract} 

\pacs{73.??}

\noindent
The fact that quantum systems may exhibit nontrivial topological
effects is known for long \cite{topology}. Recent interest to vortices
has been focused mostly at superconducting systems described by the
Ginzburg--Landau equation \cite{GL}. However, vortices have been
also observed in pure quantum mechanics, specifically in numerical
analysis of various models of mesoscopic electron transport
\cite{vortices} where they may arise even without an applied magnetic
field.

The vortical behavior in quantum mechanical scattering is closely
related with the wavefunction phase.  Writing $\,\psi(\vec
r)=\sqrt{\rho(\vec r)}\, e^{i\phi(\vec r)}\,$, we can express the
probability current as $\,\vec j(\vec r)= \rho(\vec r)\vec\nabla
\phi(\vec r)\,$ \cite{units}. In a region where external forces are
absent, the integral of $\,\vec j(\vec r)\,$ over a closed loop can
be nonzero only if it encircles a singularity in which the phase
$\,\phi\,$ is ambiguous. However, solutions to the stationary
Schr\"odinger equation are smooth functions, so only such
singularities are zeros of $\,\psi\,$.

The existing results mentioned above are usually concerned with
two--dimesional systems where the vortices are planar and centered
around nodal points. The aim of the present letter is to show how the
probability flow vortices look like in three dimensions, where nodal
sets of the scattering wavefunctions are generically smooth curves.
While the examples of Ref.\cite{vortices} deal with perturbed
channels of various shapes, the presence of the boundary is in fact
not essential for the existence of vortical solutions. To illustrate
this we shall discuss scattering in $\,\R^3$. For the sake of
simplicity we are going to analyze a simple example of a particle
scattered on a finite family of point interactions, however, the
conclusions extend to a much wider family of potentials.

We shall demonstrate that scattering wavefunctions have often nodal
lines, and that the latter are of the form of {\em closed loops.} The
probability current in the vicinity of the line is locally cylindric
(tornado--shaped). Moreover, if the scattered particles are charged,
the corresponding electric current generates a magnetic field whose 
flux lines not far from the nodal lines are closed.

Let us describe the model.  The point interactions will be treated in
the standard way \cite{AGHH}. We suppose that they are finitely many
and supported by a set  $\,Y:=\{\,\vec y_j\,:\; j=1,\dots,N\,\}\,$.
In the vicinity of the point $\,\,\vec y_j\,$ any solution to the
stationary Schr\"odinger equation behaves as 
\begin{equation} \label{delta}
\psi(\vec x)\,=\, {A_j\over 4\pi|\vec r-\vec y_j|}\,+\, B_j
+\,\OO(|\vec r-\vec y_j|)\,,
\end{equation}
where the coefficients are related by $\,B_j +\,\alpha_j A_j\,=\,0\,$
and the real parameter $\,\alpha_j\,$ characterizes the interaction
``strength" \cite{ES}. 

Suppose that the incident particle momentum is parallel to the first
axis and equals $\,k\,$. For a given family of the coupling
constants, $\,\alpha:=\{\alpha_1,\dots,\alpha_N\}\,$, the scattering
wavefunction equals \cite[Sec.~II.1.5]{AGHH}
\begin{equation} \label{scattering}
\psi_{\alpha,Y}(k;\vec x)= e^{ikx}+\, 
\sum_{j,\ell=1}^N \left\lbrack \Gamma_{\alpha,Y}(k) \right\rbrack
_{j\ell}^{-1}\, e^{iky_{\ell 1}}\, {e^{ik|\vec x-\vec y_j|} \over
4\pi|\vec x-\vec y_j|} 
\end{equation}
with $\,\vec x=(x,y,z)\,,\; \vec y_j= (y_{j1},y_{j2},y_{j3})\,$, and
\begin{equation} \label{Gamma}
\Gamma_{\alpha,Y}(k)\,:=\, \left\lbrack \left( \alpha-\,{ik\over
4\pi} \right)\delta_{j\ell}\,-\, \tilde G_k\left( \vec y_j-\vec y_\ell
\right) \right\rbrack _{j,\ell=1}^N\,,
\end{equation}
where $\,\tilde G_k\,$ is the regularized Green's function,
$$
\tilde G_k(\vec x)\,=\, {e^{ik|\vec x|} \over 4\pi|\vec x|}
$$
for $\,|\vec x|\ne 0\,$ and zero otherwise.

We denote by $\,\RR,\,\II\,$ the real and imaginary part of the
function $\,\psi_{\alpha,Y}(k;\cdot)\,$, respectively; to find nodal
sets one has to solve the equations
\begin{equation} \label{nodes}
\RR(x,y,z)\,=\,0\,=\, \II(x,y,z)\,.
\end{equation}
Some conclusions about the existence and properties of the solutions
can be derived from the implicit--function theorem \cite{implicit}: 
\begin{description}
\vspace{-1.2ex}
\item{\em (i)} In the absence of the scatterers, solutions to each of
the conditions (\ref{nodes}) are planes perpendicular to the
$\,x$--axis. They can be used as ``unperturbed" solutions, since 
$$
{\partial\RR\over\partial x} \left(\left(n+{1\over 2}\right)\pi,
y,z\right) \ne 0\,, \quad
{\partial\II\over\partial x} \left(n\pi,y,z\right) \ne 0
$$
as $\,y^2\!+z^2 \to \infty\,$. Hence far enough from the scatterers
the solutions to the two conditions are locally unique.
\vspace{-1.2ex}
\item{\em (ii)} Whenever a unique solution exists, it is locally a
$\,C^{\infty}\,$ surface, since $\,\RR,\,\II\,$ are real--analytic
functions of the variables $\,x,y,z\,$. For the same reason, the
sought nodal sets are $\,C^{\infty}\,$ curves being intersections of
two surfaces, except for possible crossing points where the
solution is locally nonunique.
\vspace{-1.2ex}
\item{\em (iii)} The implicit--function theorem also implies that the
nodal lines are confined to a bounded region of the configuration
space only. Indeed, the  zero surfaces are for large
$\,y^2\!+z^2\,$ of the form
$$
x_1\,=\,\left(n+{1\over 2}\right)\pi + f_n(y,z)\,, \quad 
x_1\,=\,n\pi + g_n(y,z)\,,
$$
respectively, and the smooth functions $\,f_n,\, g_n\,$ are
$\,\OO\left( (y^2\!+z^2)^{-1/2} \right)\,$, so that there is no
intersection for $\,y^2\!+z^2$ large enough; for large $\,|n|\,$ the
argument can be extended up to the $\,x$--axis.
\vspace{-1.2ex}
\end{description}
Before we shall discuss whether the solutions do exist, let us point
out two other properties:
\begin{description}
\vspace{-1.2ex} 
\item{\em (iv)} It follows from {\em (iii)} that the nodal line, if
they exist, consist of a family of closed loops, each of them being a
$\,C^{\infty}\,$ curve. They can cross; this happens if one of the
surfaces has locally a saddle shape and the other one ---
apropriately shifted by parameter choice --- is ``flatter", i.e., has
a small enough mean curvature.
\vspace{-1.2ex}
\item{\em (v)} On the other hand, the loops cannot entagle into
nontrivial knots, because any closed loop on a smooth surface without
intersections is topologically equivalent to a circle; one cannot
``draw" a knot upon a surface.
\vspace{-1.2ex}
\end{description}

Before proceeding further, let us remark that the above argument has
two basic ingredients. One is the finite range of the interaction
which implies that far from the scattering centers the zero surfaces
of the functions $\,\RR\,$ and $\,\II\,$ cannot intersect. The other
is the analyticity of the wavefunction with respect to the
coordinates. Since this property persists if point scatterers are
replaced by finite hard obstacles or a smooth enough potential, the
conclusions are expected to apply to a much wider class of
three--dimensional scattering systems.

To show that the described scheme is not empty, let us discuss next
the case of a single $\,\delta\,$ potential of the ``strength"
$\,\alpha\,$ situated at the origin. In view of the cylindrical
symmetry, possible nodal lines are circular loops perpendicular to
the $\,x$--axis, hence we replace $\,(y,z)\,$ by $\,(y\cos\varphi,
y\sin\varphi)\,$ so $\,|\vec x|= \sqrt{x^2\!+y^2}\,$. The scattering
solution (\ref{scattering}) now acquires the form
\begin{equation} \label{1-scattering}
\psi_{\alpha,0}(k;\vec x)\,=\, e^{ikx}\,+\, {e^{ik|\vec x|} \over
(4\pi\alpha -ik)|\vec x|}\,,
\end{equation}
where the right side turns to zero if
\begin{equation} \label{1-node}
{e^{ik(|\vec x|-x)} \over 4\pi|\vec x|}\,=\,
{ik\over4\pi}\,-\,\alpha\,. 
\end{equation}
Taking the modulus, we find the distance of the ring from the origin,
\begin{equation} \label{distance}
|\vec x|\,=\, {1\over \sqrt{k^2\!+16\pi^2\alpha^2}}\,.
\end{equation}
Furthermore, comparing the real and imaginary part in (\ref{1-node})
we find
\begin{equation} \label{difference}
|\vec x|-x\,=\,-\, {1\over k}\:{\rm Arctan\,} {k\over 4\pi\alpha}\,,
\end{equation}
where the arctangent branch remains to specified. Denoting the right
side of (\ref{difference}) as $\,-\gamma\,$ and  substituting into
$\,|\vec x|^2= x^2\!+y^2\,$, we get 
\begin{equation} \label{radius}
y^2\,=\, \gamma^2-2\gamma x\,=\, -\gamma(\gamma+2|\vec x|)\,.
\end{equation}
A nontrivial nodal ring exists if the left side is positive. It is
straightforward to see from here that this happens if $\,-2|\vec
x|<\gamma <0\,$; introducing $\,\kappa:= k/4\pi\alpha\,$ we can
rewrite the last condition as 
\begin{equation} \label{ring existence}
-\, {2|\kappa|\over \sqrt{1+\kappa^2}} \,<\, {\rm Arctan\,}\kappa
\,<\,0\,. 
\end{equation}
At the edges of the interval the ring shrinks into a nodal point.
Inspecting the inequalities (\ref{ring existence}), we infer that for
$\,\alpha<0\,$ there is a unique solution given by the basic branch.
On the other hand, one has to use $\,\arctan\kappa-n\pi\,$ if
$\,\alpha>0\,$ and the solution exists provided $\,\kappa> 2.971\,$,
i.e. 
\begin{equation} \label{existence condition}
{\alpha\over k}\,<\, 2.679\,\times\, 10^{-2}\,.
\end{equation}

If there is more than one point scatterer, it is no longer possible
to find the nodal lines analytically. However, the explicit form
(\ref{scattering}) of the scattering wavefunction allows a numerical
treatment. This demonstrates not only that the nodal loops can exist
in the multicenter situation too, but also their possible crossing
described in point {\em (v)} above; changing appropriately the
parameters, one can achieve that a loop undergoes a fission process
in which it is first ``strangled" and then decays into two loops.
As an illustration we show in Fig.~1. nodal lines inside a given
rectangular ``box" corresponding to 10 scattering centers whose
positions are marked by dots. All the point interactions have the
same strength $\,\alpha=0\,$. To convey the their spatial shapes, we
do not plot the lines themselves but rather their properly lighted
tubular neighborhoods of radius $\,R=0.1\,$.


To understand the behavior of the wavefunction in the vicinity of the
nodal lines, one has to realize that due to the smoothness of the
latter, the wavefunction has locally an approximate cylindrical
symmetry. Hence if we want to expand $\,\psi_{\alpha,Y}(k;\vec x)\,$,
it is natural to place the point of interest into the origin with the
nodal line tangent to the $\,z\,$ axis, we can use again the
coordinates of the above example and write
\begin{equation} \label{local cylinder}
\psi_{\alpha,Y}(k;\vec r)\,=\, \sum_{m\ne 0} c_m
e^{im\varphi}J_m(ky)\,. 
\end{equation}
The $\,m=0\,$ term is missing because the wavefunction vanishes by
assumption at $\,y=0\,$. The phase behavior around the node is
determined by the terms with the lowest nonzero $\,|m|\,$. Suppose
that $\,c_1\,$ or $\,c_{-1}\,$ is nonzero and $\,|c_1|\ne
|c_{-1}|\,$. Up to higher--order terms, the probability current
$\,\vec j(\vec r)\,$ is then located in a plane  perpendicular to the nodal line and 
\begin{equation} \label{invariant}
\int_{\CC} \vec v(\vec r)\,d\vec r\,=\, 2\pi\, {\rm
sgn}(|c_1|-|c_{-1}|) \;; 
\end{equation}
where $\vec v$ denotes the velocity of the probability flow, $\,\vec
v(\vec r) := \vec\nabla \phi(\vec r)\,$. By analyticity this result
extends to any curve $\,\CC\,$ encircling the nodal line once as long
as it stays away of the scattering centers. It is moreover clear that
the decribed case is generic,  because it corresponds to simple zeros
of the wavefunction. Higher winding numbers may appear only if there
is $\,m_0>0\,$ such that $\,c_m=0\,$ for
$\,m=-m_0,-m_0\!+1,\dots,m_0\,$.   

To illustrate these considerations we choose a part of the
``lowest" nodal loop of Fig.~1. To show that this is indeed the
$\,|m|=1\,$ case, we plot in Fig.~2 the wavefunction phase modulo
$\,2\pi\,$ in the planes perpendicular to the loop. A single cut is
clearly visible. In Fig.~3 we plot currents vectors at a tubular
surface centered at the nodal line.  



Let us ask whether the existence of ``tornados" can have observable
consequences. We have remarked that scattering of charged particles
may produce in this way closed magnetic flux lines. Consider
therefore a scattering of electrons and heavy ions in the presence of
fixed obstacles. If the ion--obstacle configuration is properly
chosen, the electrons should exhibit interference behavior of the AB
type.

In conclusion, we have shown here analytically and demonstrated
numerically that the probability flux in three--dimensional
scattering can exhibit vortical behavior around nodal lines of the
wavefunction. The latter are generically closed smooth curves which
may intersect but do not entangle into knots, with the probability
current velocity integral over a curve encircling the nodal line once equal to
$\,\pm 2\pi\,$.

\acknowledgments
The research has been partially supported by the Theoretical Physics 
Foundation in Slemeno and by GACR under the contract
202--0218/96--98.

\subsection*{Figure captions}

\begin{description}

\item{\bf Figure 1.} Nodal lines of the wavefunction (shown by means
of their lighted tubular neighborhoods) for scattering on 10
$\,\delta\,$ potentials of the same strength $\,\alpha=0\,$, whose
positions are marked by dots, with the incident particle momentum
$\,k=2\,$.

\item{\bf Figure 2.} Wavefunction phase around the ``lowest" loop of
the preceding picture. The wavefunction phase $\,\phi(\vec r)\,$ has
been evaluated $\,{\rm mod\,}2\pi\,$ on a tubular surface of radius
$0.1$ along the nodal line and plotted in the perpendicular planes
being scaled down by the factor $0.02$.

\item{\bf Figure 3.} The probability flow at the same tubular surface
as in Fig.~2, the observation point is changed.

\end{description}


\begin{thebibliography}{99}
\bibitem{units}
Since values of the constants are not important in the present
letter, we put $\,\hbar= 2m= 1\,$.
\end{thebibliography}
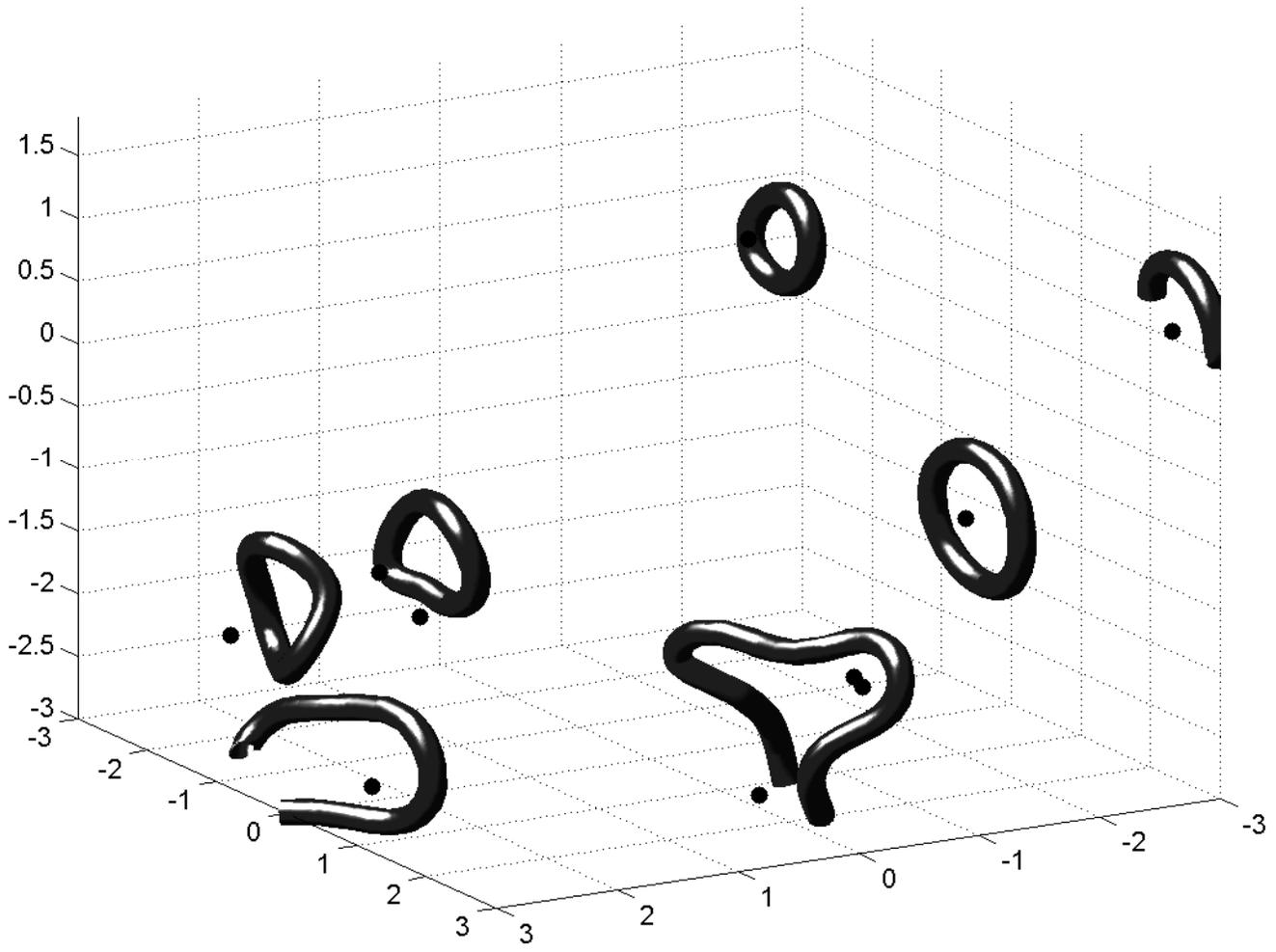
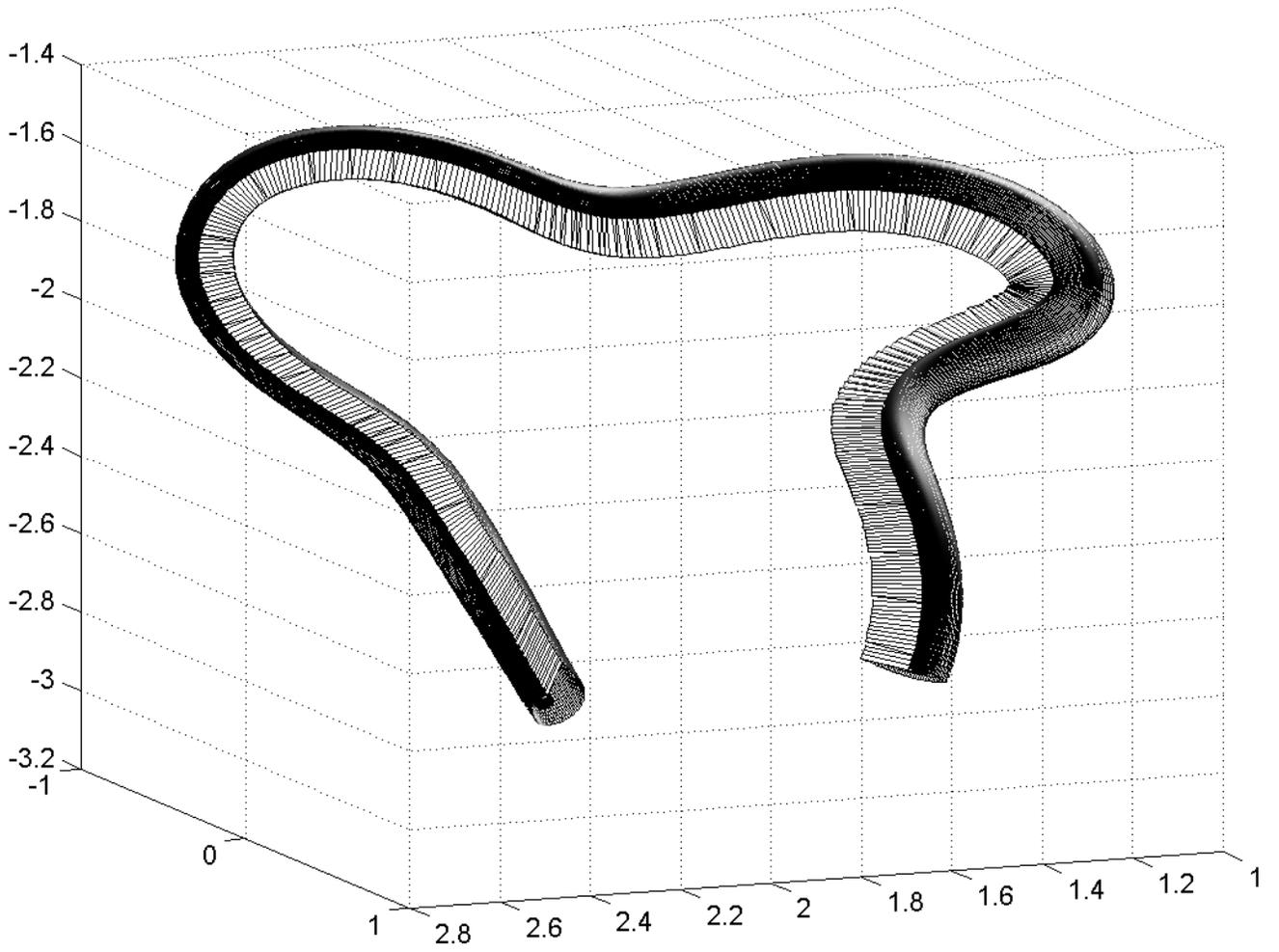
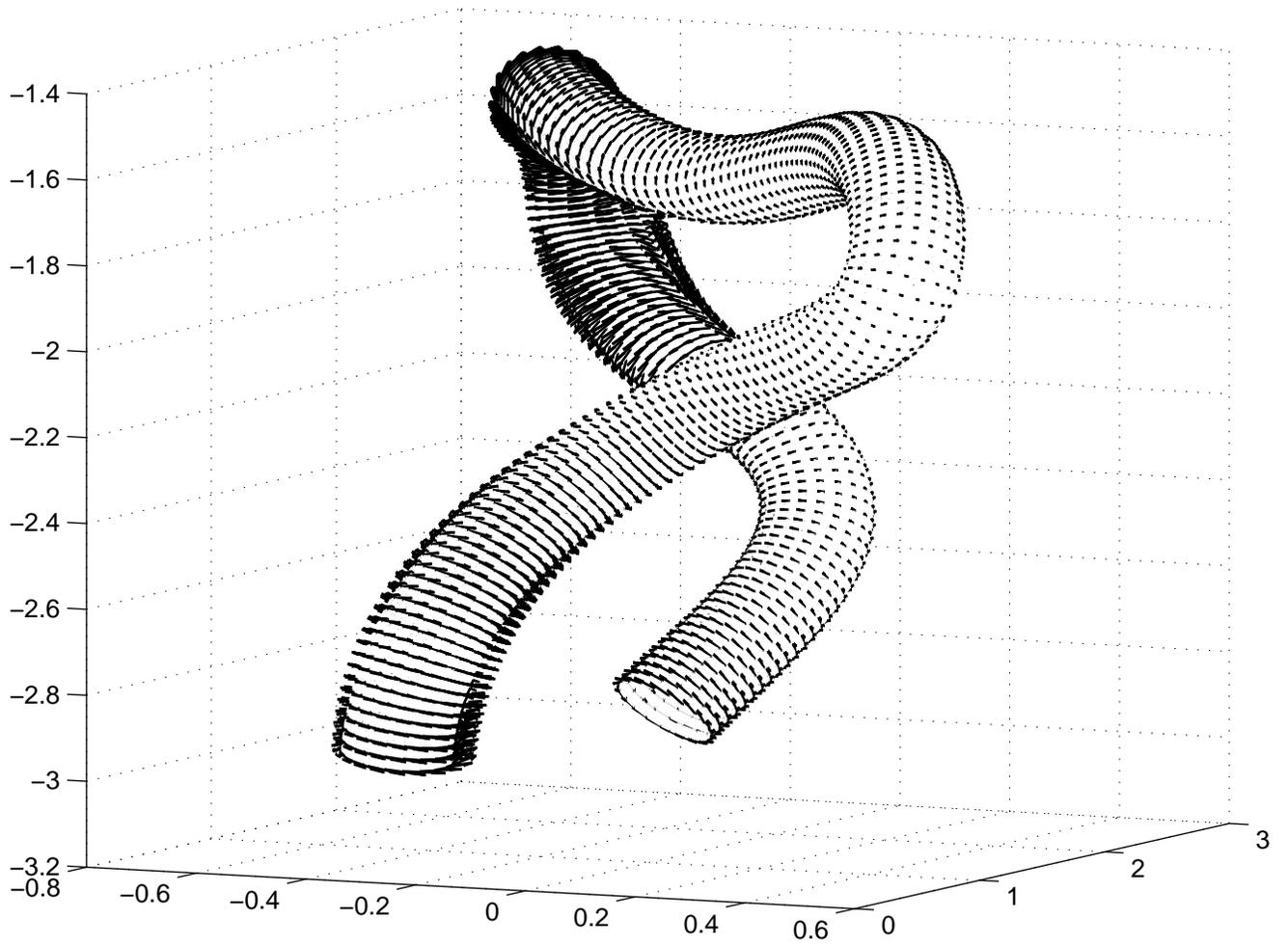
\end{document}